\documentclass[12pt]{article}
\usepackage{amsmath,amssymb,amsfonts}
\usepackage{graphicx}
\usepackage{booktabs,colortbl}

\newcommand{\be}{\begin{equation}} 
\newcommand{\ee}{\end{equation}} 
\newcommand{\bea}{\begin{eqnarray}}  
\newcommand{\eea}{\end{eqnarray}}
\newcommand{\bs}{\begin{split}} 
\newcommand{\es}{\end{split}}

\newcommand{\units}[1]{~\mathrm{#1}}

\newcommand{\scriptmath}[1]{{\scriptsize{\mbox{$#1$}}}}

\newcommand{\hc}{\mathrm{h.c.}} 
\newcommand{\mt}[1]{\mathrm{#1}}

\begin{document}




\begin{center}

\renewcommand{\thefootnote}{\fnsymbol{footnote}}

{\Large {\bf Exploring singlet deflection of gauge mediation
}} \\
\vspace*{0.75cm}
{\bf J.\ de Blas}\footnote{E-mail: jdeblasm@nd.edu}
and
{\bf A.\ Delgado}\footnote{E-mail: antonio.delgado@nd.edu}

\vspace{0.5cm}

Department of Physics, University of Notre Dame,\\
Notre Dame, Indiana 46556, USA

\end{center}

\vspace{0.5cm}

\begin{abstract}

\noindent We embed the Next-to Minimal Supersymmetric Standard Model into gauge mediation of supersymmetry breaking and study the phenomenology of scenarios where the gauge-mediation contributions to soft parameters are deflected by superpotential interactions of the gauge singlet with the messenger fields and the Higgs doublets. This kind of models provides a satisfactory solution to the $\mu$-$b_\mu$ problem of gauge mediation, compatible with the adequate pattern of electroweak symmetry breaking and a realistic spectrum with supersymmetric partners at the TeV scale without requiring a significant fine tuning.

\end{abstract}

\vspace{0.25cm}

\renewcommand{\thefootnote}{\arabic{footnote}}
\setcounter{footnote}{0}


\section{Introduction: The $\mu$-$b_\mu$ problem in gauge mediation of supersymmetry breaking}
\label{section_Intro}

Despite the many virtues that make theories with supersymmetry (SUSY) one of the most appealing candidates to extend the Standard Model (SM) at high energies, these are not absent of their own problems. The most obvious one is to explain why nature is not supersymmetric at low energies, i.e., to provide a satisfactory mechanism of SUSY breaking. This problem is intimately related with the predictive power of SUSY. Indeed, even for the minimal supersymmetric extension of the SM, the MSSM, with softly broken SUSY there are far too many free parameters to be determined experimentally before we can start making general predictions. One expects that once we know the details of the sector responsible of SUSY breaking all those parameters can be expressed in term of (hopefully) only a few (more fundamental) quantities. Models with Gauge Mediation of Supersymmetry Breaking (GMSB) \cite{Dine:1981gu,Nappi:1982hm,AlvarezGaume:1981wy,Dine:1993yw,Dine:1994vc,Dine:1995ag,Giudice:1998bp}
 offer a complete pattern of soft interactions computable within a renormalizable framework in terms of very few parameters. In GMSB one assumes that the MSSM fields feel supersymmetry breaking through the SM gauge interactions of a set of messenger fields, $\Phi$, directly coupled to a hidden sector where SUSY breaking occurs. The latter is parametrized by a gauge-singlet chiral auxiliary superfield, $X$, whose vacuum expectation value (vev) $\left<X\right>=M+\theta^2 F$ is assumed to be the only source of SUSY breaking. Through superpotential interactions $W\supset X\Phi^2$, $\left<X\right>$ generates a splitting $\sim \sqrt{F}$ between the masses (of order $M$) of the scalar and fermionic components of the messenger fields. Those effects are translated into the visible sector through radiative corrections. In this way, gaugino masses are generated at the one-loop order while scalar masses squared receive contributions only through two-loop graphs:
\be
M_{\lambda_a} \sim \frac{g_a^2}{16\pi^2}\frac{F}{M},~~~~~~~~m_{\phi_i}^2\sim\sum_a\frac{g_a^4}{256\pi^4}\frac{F^2}{M^2}\sim M_\lambda^2.
\ee
Notice that, since gauge interactions are flavor blind, the gauge-mediation contributions to the scalar masses for the sfermions are family universal. Thus, GMSB provides a scenario where one can easily avoid introducing new sources of flavor violation in the low-energy theory. Together with its predictive power, this is another of the reasons why gauge mediation has become one of the most popular mechanisms of SUSY breaking.

Another problem common to all supersymmetric extensions of the SM is the origin of $\mu$, the Higgs bilinear term in the superpotential, $W\supset \mu \left(H_u\cdot H_d\right)$ (with ``$\cdot$'' being the $SU\left(2\right)_L$ product). For phenomenological reasons $\mu$ must be of the order of the weak scale and therefore it should not be present in the limit of exact supersymmetry, for it would be naturally of the order of the Planck or another large fundamental scale. Thus, $\mu$ must be generated upon SUSY breaking. The problem in GMSB is that the same interactions generating $\mu$ also generate $b_\mu$, the corresponding Higgs bilinear soft term in the scalar potential, $V\supset b_\mu \left(H_u\cdot H_d\right)+\hc$, in a way such that~\cite{Dvali:1996cu}
\be
\frac{b_\mu}{\mu}\sim \frac{F}{M}.
\label{mubmu}
\ee
Since $F/M\gtrsim 10$ - $100\units{TeV}$, in order to generate adequate masses for the supersymmetric particles, Eq. (\ref{mubmu}) requires an unnatural fine tuning to reproduce electroweak symmetry breaking (EWSB).

A simple solution to this naturalness problem can be achieved within the Next-to-Minimal Supersymmetric Standard Model (NMSSM). A bare $\mu$ term is forbidden in the NMSSM by imposing a discreet $\mathbb{Z}_3$ symmetry, broken only at the electroweak scale.\footnote{The introduction of such a symmetry has been criticized because of the potential cosmological problems due to the formation of stable domain walls. This kind of discussion, however, goes beyond the scope of this paper. One possibility for avoiding this problem can be found in \cite{Panagiotakopoulos:1998yw}.}  An effective $\mu$ parameter can then be generated at low energies, by means of the vev of the new gauge-singlet chiral field, $S$, with superpotential couplings
\be
W\supset\lambda S \left(H_u\cdot H_d\right)-\frac{\kappa}{3}S^3.
\label{NMSSM ext}
\ee
Thus, the low energy dynamics generate $\mu^\mathrm{Eff}=\lambda \left<S\right>$. On the other hand, we have $b_\mu^\mathrm{Eff}=a_\lambda\left<S\right>+\lambda\kappa\left<S\right>^2$, with $a_\lambda$ the Higgs-singlet trilinear soft coupling in the scalar potential: $V \supset a_\lambda S \left(H_u\cdot H_d\right)+\hc$. Since the leading gauge-mediation contributions to $a$ terms arise at the two-loop level, and then are highly suppressed, $b_\mu^\mathrm{Eff}\sim \lambda\kappa\left<S\right>^2$ up to renormalization group (RG) running effects. Therefore, $b_\mu^\mt{Eff}/\mu^\mt{Eff}\sim \kappa \left<S\right>$, avoiding the hierarchy in (\ref{mubmu}). Note that a sizable value of the cubic coupling $\kappa$ is required to avoid the presence of too light pseudoscalars in the theory, for it explicitly breaks the global Peccei-Quinn symmetry of the NMSSM.

This minimal scenario with GMSB, however, is known to have problems to attain the correct EWSB and, at the same time, generate a phenomenologically viable spectrum~\cite{deGouvea:1997cx}. In particular, this requires to generate a large enough negative mass squared for the singlet, $m_S^2$, or large $a$ terms for the superpotential interactions in Eq.~(\ref{NMSSM ext}), $a_\lambda$ and $a_\kappa$. Neither of these can be obtained within pure GMSB since $S$ carries no gauge quantum numbers and, as explained above, $a$ terms are generically small in gauge mediation.

In Ref.~\cite{Giudice:1997ni} it was proposed that if, apart from gauge interactions, SUSY breaking is communicated also by means of direct superpotential interactions of the messengers with the NMSSM singlet $S$, negative values for $m_S^2$ and non-negligible contributions to $a_\lambda$ and $a_\kappa$ can be obtained. The phenomenological implications of such scenario were explored in \cite{Delgado:2007rz}, finding that EWSB is then compatible with a realistic spectrum. This scenario thus provides a viable solution to the $\mu$-$b_\mu$ problem of gauge mediation.\footnote{In this case $a_\lambda\sim \frac{\lambda}{16 \pi^2} \frac{F}{M}$ so, because of the loop factor, we can still avoid (\ref{mubmu}).} An alternative scenario where the singlet, the messengers and the Higgs doublets are coupled was discussed in \cite{Chacko:2001km}.  Again, this provides sizable contributions to the soft $a$ terms and $m_S^2$, but the latter require $\left|\kappa\right|\sim 1$ at the messenger scale to obtain a negative value. Indeed, the characteristic feature in both scenarios is the use of singlet interactions to generate the required extra contributions to the soft terms for $S$, that then deviate from the standard gauge mediation scenario. Hence the name of singlet deflection of gauge mediation.

As explained in \cite{Delgado:2007rz}, one of the main phenomenological difficulties of the scenario with messengers coupled only to $S$ is to obtain regions in the parameter space where the lightest CP-even Higgs boson, $H^1$, passes the direct LEP 2 bound of 114 GeV~\cite{Barate:2003sz}. This has been one of the major problems of supersymmetric extensions of the SM, where the Higgs is naturally very light at tree level and one relies in large contributions from radiative corrections to lift the leading order value above the LEP 2 bound. In general, attaining a Higgs with masses above 114 GeV within this model required to assume both a large SUSY-breaking scale $F/M$ and a large messenger scale $M$ in order for top and stop loops to give such large radiative corrections to $m_{H^1}$. This lead to a heavy spectrum where, for instance, stops were required to have masses around 2 TeV. As we will illustrate, one can lower the mass scale of that model and still find regions in the parameter space where all experimental restrictions are passed, but such regions are quite small thus requiring a significant fine tuning. 

Motivated by this tension between obtaining a relatively light spectrum and still having large allowed regions in the parameter space, in this paper we combine the ideas of \cite{Delgado:2007rz} and \cite{Chacko:2001km} to explore the scenario where deflection of gauge mediation is due to general interactions involving the singlet and the messenger fields. We show that, despite this more general scenario introduces several parameters with respect to \cite{Delgado:2007rz}, only one of them suffices to greatly increase the size of the allowed regions in the parameter space. Thus, the fine tuning problem for low SUSY-breaking scales is removed. In particular, this allows a TeV scale spectrum which may be eventually tested by the LHC.

The paper is organized as follows. In the next section we present the model and provide the soft SUSY-breaking terms in the effective theory below the messengers mass threshold. Section 3 explains the method employed in our phenomenological analysis of the model. We scan the parameter space searching for regions where EWSB can be reproduced. The resulting spectrum for those regions is detailed in section 4. Finally we present our conclusions.


\section{Description of the model}
\label{sec:Model}

We use the following conventions in writing the superpotential couplings and soft $a$ terms of the $\mathbb{Z}_3$-symmetric NMSSM: 
\be
W=-y_t~u_3^c \left(H_u \cdot q_3\right) +y_b~d_3^c \left(H_d \cdot q_3\right)+y_\tau~e_3^c \left(H_d \cdot l_3\right) +\lambda~S \left(H_u\cdot H_d\right)-\frac{\kappa}{3}~S^3,
\label{NMSSMW}
\ee
\be
{\cal L}_\mathrm{soft}\supset a_t~u_3^c\left(H_u \cdot q_3\right) -a_b~d_3^c\left(H_d \cdot q_3\right)-a_\tau~e_3^c\left(H_d \cdot l_3\right) -a_\lambda~S \left(H_u\cdot H_d\right)+\frac{a_\kappa}{3}~S^3+\hc,
\ee
where, as usual, we have neglected Yukawa couplings other than those for the third SM family.

Following \cite{Delgado:2007rz}, above the messenger scale, $M$, we will assume $n=2$ pairs of messenger fields $\Phi_i$ and $\bar{\Phi}_i$, $i=1,2$, transforming as a 5 and a $\bar{5}$ of $SU\left(5\right)$, respectively. These fields decompose into a triplet and a doublet under $SU\left(3\right)_c$ and $SU\left(2\right)_L$, respectively: $\Phi=\left(\Phi^T,\Phi^D\right)$ and couple to the spurion field, $X=M+\theta^2 F$, responsible of SUSY breaking. This is considered only as a background nondynamical field. Apart from gauge interactions the messenger fields can also have direct superpotential couplings with the NMSSM fields, thus altering the gauge-mediation pattern of soft SUSY-breaking contributions. We consider the following superpotential couplings for the messenger fields at $M$
\be
\begin{split}
W_\mt{mess}&=X\sum_{i=1}^{n=2}\left(\kappa_i^D \bar{\Phi}^D_i\Phi^D_i+\kappa_i^T \bar{\Phi}^T_i\Phi^T_i\right)+\\
&+S\left(\xi_D \bar{\Phi}^D_1\Phi^D_2+\xi_T \bar{\Phi}^T_1\Phi^T_2+\xi_{H_u} \bar{\Phi}_1^D H_u+\xi_{H_d} \left(H_d\cdot \Phi_2^D\right)\right).
\end{split}
\label{MessW}
\ee
The structure of the above superpotential can be explained under the following asumptions. As stressed in \cite{Delgado:2007rz}, the structure of the singlet-messenger couplings can be explained by extending the discrete $\mathbb{Z}_3$ symmetry of the NMSSM with $\mathbb{Z}_3[\Phi_1]=\mathbb{Z}_3[\bar{\Phi}_2]=-1/3$, $\mathbb{Z}_3[\Phi_2]=\mathbb{Z}_3[\bar{\Phi}_1]=1/3~(=\mathbb{Z}_3[S]=\mathbb{Z}_3[H_u]=\mathbb{Z}_3[H_d])$, $\mathbb{Z}_3[X]=0$. Let us note that the couplings in (\ref{MessW}) generate a kinetic $\bar{\Phi}_1^D$-$H_d$ and $\Phi_2^D$-$H_u$ mixing at the one-loop level. Indeed, we have the following contributions to the off-diagonal elements of the matrix of anomalous dimensions $\gamma$~\footnote{~$\gamma_i^j\equiv d \log{Z_i^j}/d\log{Q}$, with $Z_i^j$ the corresponding wavefunction renormalization.} above the messenger scale:
\be
\gamma_{\bar{\Phi}_1^D}^{H_d}=-\frac{1}{8\pi^2}\left(\lambda \xi_{H_u}+\xi_D\xi_{H_d}\right),
\label{WFR1}
\ee
\be
\gamma_{\Phi_2^D}^{H_u}=-\frac{1}{8\pi^2}\left(\lambda \xi_{H_d}+\xi_D\xi_{H_u}\right).
\label{WFR2}
\ee
Thus, even if one of the couplings $\xi_D$, $\xi_{H_u}$ or $\xi_{H_d}$ vanishes at a given scale $Q\geq M$, a nonzero value is generated through renormalization above the messenger scale. This is not in contradiction with the nonrenormalization theorem, since the effect comes from the wavefunction renormalization. The above $\mathbb{Z}_3$ symmetry still allows for the possibility of including terms like $X \bar{\Phi}_2^D H_u $ and $X \left(H_d \cdot \Phi_1^D\right)$. Upon SUSY breaking, these introduce a mixing between the NMSSM Higgs fields and the doublet components of the messenger fields. Such mixing should be diagonalized in order to obtain the physical Higgses and messenger doublets, giving rise to additional interactions in the superpotential between the physical messengers and the NMSSM fields \cite{Chacko:2001km}. Note that, because of the mixing induced by (\ref{WFR1}) and (\ref{WFR2}), RG effects above $M$ would also generate the above-mentioned terms, even if we assume they vanish at a given scale. In this regard, for simplicity and to avoid the proliferation of new parameters controlling the contributions to soft terms we will assume there is no such mixing at the messenger scale. Thus, in the above superpotential $H_{d,u}$ describe the physical Higgses and these correspond to the NMSSM ones. Genuine trilinear couplings between the doublet components of the messengers and the quark/squark or lepton/slepton chiral fields could be also present:
\be
\Delta W_\mt{mess}=- \xi_u^{ij}~u_i^c  \left(\Phi_2^D \cdot q_j\right) +\xi_d^{ij}~d_i^c~\bar{\Phi}_1^D  q_j+\xi_e^{ij}~e_i^c~\bar{\Phi}_1^D l_j.
\label{MessW2}
\ee
Should they be present, a rather unnatural alignment of such couplings with the Yukawa ones would be required in order to avoid introducing extra sources of flavor violation. On the other hand, this kind of interactions can be forbidden within extra dimensional models by locating in different branes the MSSM quarks and leptons, and the hidden and messenger fields. Here, again for simplicity, we will just assume that the couplings in (\ref{MessW2}) vanish at the messenger scale. In this case, for energy scales $Q>M$ the wavefunction renormalizations (\ref{WFR1}) and (\ref{WFR2}) would generate nonzero values only for the couplings with the third family~\footnote{In turn, these feed (\ref{WFR1}) and (\ref{WFR2}) with extra contributions.} ($\xi_t\equiv \xi_u^{33}$, $\xi_b\equiv\xi_d^{33}$ and $\xi_\tau\equiv \xi_e^{33}$). Note also that in (\ref{MessW}) one can write similar couplings replacing $H_d$ by the left-handed lepton chiral fields $l_i$. These can be avoided by simply requiring $R$-parity conservation. Finally, just to mention that, as also stressed in \cite{Delgado:2007rz}, the two messengers are required in order to avoid kinetic mixing between $X$ and $S$, which could destabilize the weak scale. The structure of the minimal superpotential in Eq. (\ref{MessW}) is preserved if we include extra messenger fields, as long as they come in pairs and we enlarge consequently the assignments of the $\mathbb{Z}_3$ symmetry.

Upon integrating the messenger fields out, the couplings in Eq. (\ref{MessW}) generate the following contributions to the soft parameters at $Q=M$. These can be derived using the general method described in \cite{Giudice:1997ni,ArkaniHamed:1998kj}. For the scalar soft masses squared the contributions at the leading order in $F^2/M^2$ arise at two loops. These deviate from the gauge-mediation contributions by terms controlled by both the new interactions in (\ref{MessW}) and the standard superpotential couplings for each particle:

\be
\begin{split}
m_{q_3}^2=&\frac{1}{256 \pi^4}\left(\frac{g_1^4}{15}+3g_2^4+\frac{16}{3}g_3^4-y_b^2\xi_{H_d}^2-y_t^2\xi_{H_u}^2\right)\frac{F^2}{M^2},\\
m_{u^c_3}^2=&\frac{1}{256 \pi^4}\left(\frac{16}{15}g_1^4+\frac{16}{3}g_3^4-2y_t^2\xi_{H_u}^2\right)\frac{F^2}{M^2},\\
m_{d^c_3}^2=&\frac{1}{256 \pi^4}\left(\frac{4}{15}g_1^4+\frac{16}{3}g_3^4-2y_b^2\xi_{H_d}^2\right)\frac{F^2}{M^2},\\
m_{l_3}^2=&\frac{1}{256 \pi^4}\left(\frac{3}{5}g_1^4+3 g_2^4-y_\tau^2\xi_{H_d}^2\right)\frac{F^2}{M^2},\\
m_{e^c_3}^2=&\frac{1}{256 \pi^4}\left(\frac{12}{5}g_1^4-2y_\tau^2\xi_{H_d}^2\right)\frac{F^2}{M^2},
\end{split}
\label{sf3mass}
\ee

\be
\begin{split}
m_{H_d}^2=&\frac{1}{256 \pi^4}\left[\frac{3}{5}g_1^4+3g_2^4-\xi_{H_d}^2\left(\frac{3}{5}g_1^2+3g_2^2-2\kappa^2-4\xi_{H_d}^2-4\xi_D^2-3\xi_T^2-2\xi_{H_u}^2\right)-\right.\\
-&\left.2\lambda^2\left(\xi_D^2+\frac{3}{2}\xi_T^2+2\xi_{H_u}^2\right)\right]\frac{F^2}{M^2},\\
m_{H_u}^2=&\frac{1}{256 \pi^4}\left[\frac{3}{5}g_1^4+3g_2^4-\xi_{H_u}^2\left(\frac{3}{5}g_1^2+3g_2^2-2\kappa^2-4\xi_{H_u}^2-4\xi_D^2-3\xi_T^2-2\xi_{H_d}^2\right)-\right.\\
-&\left.2\lambda^2\left(\xi_D^2+\frac{3}{2}\xi_T^2+2\xi_{H_d}^2\right)\right]\frac{F^2}{M^2},\\
m_{S}^2=&\frac{1}{256 \pi^4}\left[-\frac{6}{5}g_1^2\left(\xi_{H_d}^2+\xi_{H_u}^2+\xi_D^2+\frac{2}{3}\xi_T^2\right)-6g_2^2\left(\xi_{H_d}^2+\xi_{H_u}^2+\xi_D^2\right)-16g_3^2\xi_T^2+\right.\\
+&8\xi_{H_u}^2\left(\frac{3}{4}y_t^2-\kappa^2+\xi_{H_u}^2+2\xi_D^2+\frac{3}{2}\xi_T^2\right)+8\xi_{H_d}^2\left(\frac{3}{4}y_b^2+\frac{1}{4}y_\tau^2-\kappa^2+\xi_{H_d}^2+2\xi_D^2+\frac{3}{2}\xi_T^2\right)+\\
+&\left.8\xi_{H_u}^2\xi_{H_d}^2+8\xi_D^2\left(\xi_D^2-\kappa^2\right)+3\xi_T^2\left(5\xi_T^2-4\kappa^2\right)+12\xi_D^2\xi_T^2+8\lambda\xi_{H_d}\xi_{H_u}\xi_D\right]\frac{F^2}{M^2},
\end{split}
\label{HSmass}
\ee
where $g_{1,2,3}$ are the SM gauge couplings with $g_1\equiv \sqrt{5/3}g^\prime$ and the hypercharge defined by $Q=T_3+Y$. These are the leading contributions provided $F/M^2\ll 1/4\pi$. Otherwise, one-loop corrections generated at order $F^4/M^6$ from nontrivial terms in the effective K{\"a}hler potential can be comparable. Here we will concentrate in the previous regime, where such one-loop contributions can be safely neglected. For the first and second sfermion families only the gauge-mediation contributions are considered since we neglect the corresponding Yukawa couplings.
The soft $a$ terms, which are negligible in GMSB, now receive the leading contributions at one loop:
\be
\begin{split}
a_t=&-\frac{y_t \xi_{H_u}^2}{16\pi^2}\frac{F}{M},\\
a_b=&-\frac{y_b \xi_{H_d}^2}{16\pi^2}\frac{F}{M},\\
a_\tau=&-\frac{y_\tau \xi_{H_d}^2}{16\pi^2}\frac{F}{M},\\
a_\lambda=&-\frac{1}{16\pi^2}\left[\lambda\left(4\xi_{H_d}^2+4\xi_{H_u}^2+2\xi^2_D+3\xi^2_T\right)+2\xi_{H_d}\xi_{H_u}\xi_D\right]\frac{F}{M},\\
a_\kappa=&-3\frac{\kappa}{16\pi^2}\left(2\xi_{H_d}^2+2\xi_{H_u}^2+2\xi_D^2+3\xi_T^2\right)\frac{F}{M}.
\end{split}
\label{aterms}
\ee
Thus, this model has enough freedom to provide for large enough negative contributions to $m_S^2$, $a_\lambda$ and $a_\kappa$, as required to make the correct EWSB compatible with a realistic spectrum. In particular, compared to the scenario with only $S \bar{\Phi}^D_1 H_u$ couplings in \cite{Chacko:2001km}, the singlet-messenger interactions can turn $m_S^2$ negative without requiring large values of $|\kappa|\sim 1$. On the other hand, it is also possible to generate a large $a_t$ and therefore a large stop mixing. This enhances the size of the radiative top/stop one-loop corrections to $m_{H^1}$. Thus, it should be possible to lift the Higgs mass prediction above the direct LEP 2 bound with relaxed  (not too large) values of $F/M$ and $M$, compared to the case of the model without $H_u$ couplings to the messenger fields.

Finally, we have the gaugino masses. These are given by the pure GMSB contribution. At one loop,
\be
M_{\lambda_a}=\frac{g_a^2}{8\pi^2}\frac{F}{M},~~a=1,2,3,
\ee
for $n=2$ messenger fields.


\section{Electroweak symmetry breaking and the spectrum calculation}

Using the model described in the previous section we have performed a scan in the parameter space looking for those regions where a satisfactory EWSB is possible together with a phenomenologically acceptable spectrum. In what follows, and prior to show the results of this analysis we sketch the method employed in the scan. This largely follows that of Ref. \cite{Delgado:2007rz} so we refer to that reference for more details.

The model has eight unknown input parameters: the superpotential couplings $\lambda$, $\kappa$, $\xi_D$, $\xi_T$, $\xi_{H_d}$ and  $\xi_{H_u}$, and the SUSY breaking and messenger scales $F/M$ and $M$. The values for the SM parameters $g_{1,2,3}$, $y_{t,b,\tau}$ and the electroweak scale, $v^2\equiv \left<H_d\right>^2+\left<H_u\right>^2\approx \left(174\units{GeV}\right)^2$, can be determined from the experimental values of $G_F$, $M_Z$, $\alpha_\mt{em}$, $\alpha_\mt{S}\left(M_Z^2\right)$, and the known fermion masses \cite{Nakamura:2010zzi} (for the top mass we use $m_t=173.3 \units{GeV}$ \cite{:1900yx} and also take into account  one-loop QCD corrections in extracting the corresponding Yukawa coupling). Using the RG equations to evolve all these parameters from the scales where they are defined up to the messenger scale we can compute the values of the SUSY-breaking soft terms using Eqs. (\ref{sf3mass}), (\ref{HSmass}) and (\ref{aterms}). Then we must use again the RG to run all the couplings down to the scale where we minimize the scalar potential
\be
\begin{split}
V&=m_{H_u}^2 \left|H_u^0\right|^2+m_{H_d}^2 \left|H_d^0\right|^2+m_{S}^2 \left|S\right|^2+\left(-a_\lambda S H_u^0 H_d^0-\frac{a_\kappa}{3}S^3+\hc\right)+\\
&+\left|\lambda\right|^2\left|S\right|^2\left(\left|H_u^0\right|^2+\left|H_d^0\right|^2\right)+\left|\lambda H_u^0 H_d^0+\kappa S^2\right|^2+\frac{\frac{3}{5}g^2_1+g^2_2}{8}\left(\left|H_d^0\right|^2-\left|H_u^0\right|^2\right)^2.
\end{split}
\label{Vscal}
\ee
The dominant ${\cal O}\left(y_t^4,y_b^4\right)$ one-loop radiative corrections in the effective scalar potential are taken into account by replacing $V\rightarrow V+\Delta V$, where $\Delta V$ reads in the $\overline{\mt{DR}^\prime}$ scheme \cite{Barger:2006dh,Martin:2001vx}: 
\be
\Delta V=\frac{3}{32\pi^2}\sum_{f=t,b}\left[\sum_{i=1}^2 \overline{m}_{\tilde{f}_i}^4\left(\log{\frac{\overline{m}_{\tilde{f}_i}^2}{Q^2}}-\frac 32 \right)-2\overline{m}_{f}^4\left(\log{\frac{\overline{m}_{f}^2}{Q^2}}-\frac 32\right)\right],
\ee
with $\overline{m}_f$, $\overline{m}_{\tilde{f}_i}$ the field-dependent fermion/sfermion masses.

In order to minimize the (leading) ${\cal O}\left(y_t^4\right)$ corrections, as in \cite{Delgado:2007rz} we choose to perform this last step at a matching scale given by the geometric average of the stop masses, $M_\mathrm{match}=\sqrt{m_{\tilde{t}_1}m_{\tilde{t}_2}}$. This is also the scale where we specify the value for $\lambda$. On the other hand, $\kappa$ is determined in the minimization from the extremal conditions of the potential at the electroweak vacuum, together with  $\tan{\beta}$ and $\left<S\right>$. Thus, $\kappa$ is actually an output, which reduces the number of unknown parameters  by one.

We will require in our analysis that the gauge and superpotential couplings remain perturbative up to the grand unification scale, $M_\mt{GUT}$, defined as the scale such that $g_1(M_\mt{GUT})=g_2(M_\mt{GUT})$. For simplicity we will assume that the singlet-messenger interactions unify at that scale, at a common value $\xi_U\equiv \xi_D,\xi_T(M_\mt{GUT})$. This removes another parameter from the list, leaving finally with a total of six unknown inputs. The $\xi_{H_u}$ and $\xi_{H_d}$ couplings are also specified at $M_\mt{GUT}$. For completeness, the corresponding RG equations for the region $M\leq Q < M_\mt{GUT}$ are given in the appendix. Recall that, despite we assume that the messenger couplings to quark and lepton chiral fields vanish at $M$, nonzero values for $\xi_t$, $\xi_b$ and $\xi_\tau$ will be generated above the messenger scale. We take their effect on the running into account. Finally note that since the boundary conditions for the different parameters are specified at different scales and also depend on the results of the minimization, after solving the RG equations and minimizing the Higgs potential we need to adjust $\tan{\beta}$, $\left<S\right>$ and $\kappa$ and iterate until the procedure converges.

Once we obtain a set of parameters that allows to reproduce the correct EWSB we make sure that the resulting spectrum passes all existing experimental bounds~\cite{Nakamura:2010zzi}. Although such bounds depend on a variety of assumptions and may not directly apply in some cases, we prefer to be conservative in the cuts we apply prior to accept a given point in the scan. In particular, we demand that the lightest neutralino is above $50\units{GeV}$ or $120 \units{GeV}$ depending on whether it behaves as an stable or unstable particle, respectively. For the lightest chargino we also impose the cut in $120\units{GeV}$. Current limits for sfermions are different for sleptons and squarks and in the latter case also distinguish between generic squarks or the third family ones. For the sleptons and the third family of squarks we impose a common bound $m_{\tilde{l_i},{\tilde t_i},{\tilde b_i}}>120\units{GeV}$. Limits for generic squarks are significantly stronger than those of stops and sbottoms. Here we use the latest experimental results on SUSY searches at the LHC \cite{Khachatryan:2011tk,daCosta:2011hh,Collaboration:2011qk}. These correspond to the minimal supergravity/constrained MSSM parameter space and impose bounds on generic squarks and gluino masses around $700$-$800\units{GeV}$, depending on the difference between $m_{\tilde{q}}$ and $M_{\tilde g}$. For the first two families of squarks and the gluino mass we thus require $m_{\tilde q},M_{\tilde g}>700\units{GeV}$. For the Higgs masses we make use of the existing bounds from LEP 2 searches though, as we will see, in practice only the limit on a SM-like Higgs mass of $114\units{GeV}$ applies.

The mass matrices for the sfermions, neutralinos, charginos and CP even and odd Higgses can be found in the literature (see for instance \cite{King:1995vk,Ellwanger:2009dp}) and here we only sketch the precision in our determination of the lightest CP-even Higgs mass in order to compare with the LEP 2 bound. These include the dominant ${\cal O}\left(y_{t,b}^4\right)$ one-loop corrections \cite{King:1995vk} as well as the two-loop leading logarithmic corrections of ${\cal O}\left(y_{t,b}^4~\!g_3^2,y_{t,b}^6\right)$ \cite{Ellwanger:2009dp}. We also include the one-loop leading  logarithmic corrections of ${\cal O}\left(y_{t,b}^2~\!g_{1,2}^2,y_{t,b}^2~\!\lambda^2\right)$ that enter in the mass matrices through the wavefunction renormalization of the scalar fields \cite{Ellwanger:2009dp}.
Finally, we also include the one-loop leading logarithmic corrections of ${\cal O}\left(\lambda^4\right)$ to the mass of the lightest CP-even Higgs boson, computed in the limit $\left<S\right>\gg v$:
\be
\Delta m_{H^1}^2=-\frac{3\lambda_\mt{Eff}^2 v^2}{4\pi^2} \log{\frac{Q^2}{m_t^2}},
\ee
with $\lambda_\mt{Eff}$ the effective quartic coupling for the lightest CP-even Higgs boson.


\section{Phenomenological results}

Here we present the results of the scan and discuss the phenomenological features of the model. We present the results in two parts. In the first one we discuss the general aspects of the spectrum obtained from a scan over all the six free parameters. In the second, in order to compare the model with the one presented in \cite{Delgado:2007rz}, we consider a submanifold of the allowed region of the parameter space where some of the parameters are fixed to specific values. Along this section we will often refer to the model in that reference, with only singlet-messenger interactions, as N-GMSB (as denoted there) while for the general scenario of singlet deflection of GMSB presented here we use S-GMSB. We discuss the improvements with respect to \cite{Delgado:2007rz}, emphasizing whether or not these are worth increasing the number of free parameters.

We have explored values for the SUSY-breaking scale $F/M$ up to $175\units{TeV}$. For $F/M \gtrsim 175 \units{TeV}$  it was shown in \cite{Delgado:2007rz} that significantly large regions in the parameter space were allowed by all experimental bounds, provided we allow for a long enough running, i.e., a large $M$ ($F/M=172\units{TeV}$ and $M=10^{10}\units{TeV}$ in the example presented there). Here we focus on whether this can be achieved with a lower scale of SUSY breaking. Regarding the messenger mass scale we take $10^{4}\units{TeV}\leq M\leq 10^{11}\units{TeV}$. We restrict to values above $10^4\units{TeV}$ to ensure $F/M^2\ll1/4\pi$, so the leading corrections to soft masses are those in Eqs. (\ref{sf3mass}) and (\ref{HSmass}). For the remaining unknown parameters, the superpotential couplings $\lambda$, $\xi_U$, $\xi_{H_u}$ and $\xi_{H_d}$, we scan both positive and negative values. This is required since they contribute to terms with no definite sign in the boundary conditions (\ref{HSmass}) and (\ref{aterms}), and in the RG equations above the messenger scale.

As stressed in \cite{Delgado:2007rz}, one of the major difficulties in the model with only singlet-messenger interactions was finding a set of parameters that gives a large enough Higgs mass so it passes the direct LEP 2 bound. The problem in that model is the absence of direct contributions to $a_t$ at the messenger scale. Thus, a nonzero $a_t$ can be generated only through the RG evolution down to $M_\mt{match}$. This results in small values of the stop mixing, which then require large stop masses, and therefore large values of $F/M$, in order to lift $m_{H^1}$ above 114 GeV through top/stop radiative corrections. As we explained in section \ref{sec:Model}, achieving a significantly large stop mixing should be no longer an issue once we consider the new interactions between the singlet, the messengers and $H_u$. Therefore, since a larger stop mixing than in \cite{Delgado:2007rz} is in general predicted, lower values of $F/M$ should be naturally allowed, yielding also to a lighter spectrum. Still, overcoming the LEP 2 bound is the major constraint in $F/M$ and $M$, but the results of the scan show that values of the SUSY-breaking scale $F/M\gtrsim 50 \units{TeV}$ are within the allowed region. (For those values of $F/M$ the extra Higgs boson masses are still effectively decoupled. Therefore, the lightest CP-even Higgs is SM-like and the $114\units{GeV}$ limit applies, instead of the less stringent bounds from direct searches of the supersymmetric Higgses.) On the other hand, we find $m_{H^1}\lesssim 123 \units{GeV}$ for the entire set of scan points. Regarding also the Higgs mass and the radiative corrections we find 
\be
1.5\lesssim\tan{\beta}\lesssim35,
\ee
with a general preference for low values. Therefore, contributions from bottom/sbottom loops to the Higgs mass can be safely neglected in general. Nevertheless, the small regions with large $\tan{\beta}$ require to include bottom/sbottom radiative corrections in the EWSB sector as explained in the previous section, for they might start to be noticeable at that point. Note also that in that case these corrections actually lower the prediction for the Higgs mass.

\begin{figure}[t]
\input{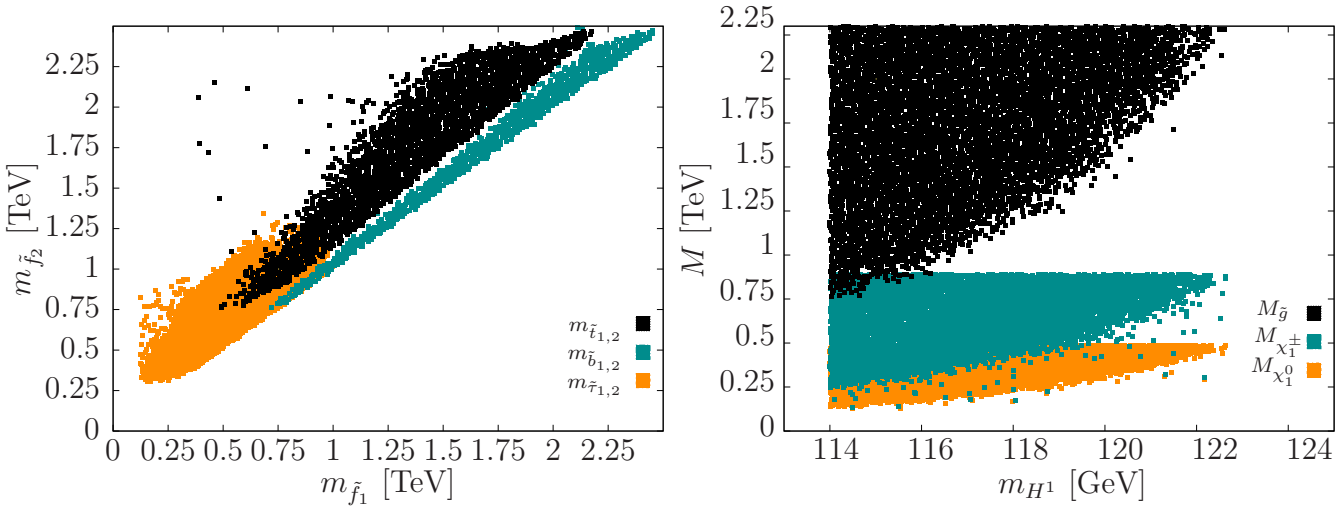}
\caption{(Left) Third family sfermion masses. (Right) Gluino and lightest neutralino and chargino masses versus the lightest CP-even Higgs mass. The sharp upper bounds correspond to the upper bound of $F/M$ in the scan. The lightest neutralino is in general the NLSP.}
\label{fig_Spectrum}
\end{figure}

On general grounds the spectrum of the model is gauge mediation like, with sparticle masses ordered according to their gauge interactions. The lightest supersymmetric particle (LSP) is the gravitino, which may become a good dark matter candidate for relatively low values of $F/M$ and $M$. As explained above such values are now easily accessible. The characteristic SUSY mass scale (roughly $\sim F/(16\pi^2 M)$) is $\gtrsim 320 \units{GeV}$. In Fig. \ref{fig_Spectrum} left we show the masses for the third family of charged sfermions. 
These are typically lighter than the first and second generations because of the effect of the Yukawa couplings. Such effects enter not only in the running but also in the boundary conditions (\ref{sf3mass}). Hence, a characteristic feature of this scenario is a larger splitting between the third and the first two families of sfermions, compared to the one with pure gauge mediation. This is more pronounced in the case of stops, since not too large values of $\tan{\beta}$ are preferred. Still, the effect is only noticeable for sizable values of the $\xi_{H_u}$ coupling. In particular, it is noteworthy that stops (and to a less extent sbottoms) below the TeV are still compatible with all the constraints. Note also the presence of some points in the region of stop masses below $500\units{GeV}$. These correspond to configurations where an approximate cancellation in the boundary condition for the third family of squark masses  is taking place (see Eq. (\ref{sf3mass})). They are, however, rather difficult to find in the scan. This may indicates that such configurations require a significant fine tuning and do not correspond to natural solutions. As in gauge mediation the lightest sfermions are the staus. Stau masses can be compared with the lightest neutralino mass, which is shown in Fig. \ref{fig_Spectrum} right, together with the gluino and the lightest chargino masses. In general, the lightest neutralino is lighter than staus, thus being the next-to-lightest supersymmetric particle (NLSP). This is the case for low $\tan \beta \lesssim 3$. For larger values there are regions where the stau can be nearly degenerate or slightly lighter than the lightest neutralino. In this case both may behave as co-NLSPs. Only for some regions with large $\tan{\beta}\gtrsim 20$ the effect of the $\tau$ Yukawa couplings can lower $m_{\tilde{\tau}_1}$ significantly below $M_{\chi^0_1}$. Regarding the lightest neutralino composition, this is in general bino like. However, we also find some regions where this can be singlino or Higgsino like. The main phenomenology of this model depends on the NLSP nature and lifetime. At particle colliders the most interesting scenarios occurs for those cases where the NLSP is stau, or it is neutralino but decays promptly within the detector. The latter occurs for $\sqrt{F}\lesssim 10^3\units{TeV}$, which is still accessible but only in a small region at the lower end of the range of allowed values of $F/M$ and $M$. For larger values the NLSP behaves like an stable particle. On the other hand, one must be aware that long-lived NLSP might pose cosmological problems for ordinary nucleosynthesis~\cite{Giudice:1998bp,Gherghetta:1998tq}. These can put additional constraints on the upper bounds of $F/M$ and $M$. For instance, for photon decays we are safe as long as $\tau_\mt{NLSP}< 10^7$ sec. For the range of values of $F/M$ explored this translates into an upper bound for the messenger scale $M\lesssim 10^{12}$-$10^{14}\units{TeV}$. Hadronic decays might impose even stronger constraints $M\lesssim 10^{9}$-$10^{11}\units{TeV}$.

Despite it is clear that the S-GMSB scenario favors heavier Higgs masses through radiative corrections than the one without the interactions involving Higgs doublets, we would like to determine quantitatively the actual improvement respect to N-GMSB, for we are introducing extra free parameters which reduce the predictive power of the model. In what follows we compare our results with those of \cite{Delgado:2007rz}. For that purpose we have considered two specific benchmark points with fixed values of $F/M$ and $M$. For the first one we have chosen $F/M=172~\units{TeV}$ and $M=10^{10}\units{TeV}$ as in the example presented in \cite{Delgado:2007rz}. As we have stressed previously, one of the major advantages of the model presented here is that it does not require of large values of $F/M$ and $M$ to pass all the constraints. Thus, we have also considered another point where both parameters take significantly lower values: $F/M=100\units{TeV}$ and $M=10^4\units{TeV}$. Let us note here that the $\xi_{H_d}$ coupling has a negligible effect over the Higgs mass prediction. Indeed, the impact of the parameter $\xi_{H_d}$ in generating a significant sbottom mixing is always irrelevant. Thus, it does not have any significant effect on the (already tiny) bottom/sbottom radiative corrections. The only relevant effect that this parameter may have is to help in the process of generating a negative $m_S^2$ and controlling the size of $a_\lambda$~\footnote{Actually, the only contribution to $a_\lambda$ that has no definite sign vanishes for $\xi_{H_d}\rightarrow0$.}. However, this r{\^o}le can be easily played by some of the other parameters. Therefore, for our purposes this coupling looks completely irrelevant. Hence, in order to emphasize the important effects in the comparison, we have gotten rid of this parameter by assuming that it vanishes at the scale $M_\mt{GUT}$, i.e., we only consider one extra parameter~\footnote{At any rate, since $\xi_{H_u}$ and $\xi_D$ are nonvanishing, small corrections from $\xi_{H_d}$ are still expected from the running down to the messenger scale (see discussion below Eq. (\ref{MessW}), and Eq. (\ref{MessCoupRGEs}) in the appendix).} compared to the N-GMSB scenario.

\begin{figure}[t]
\input{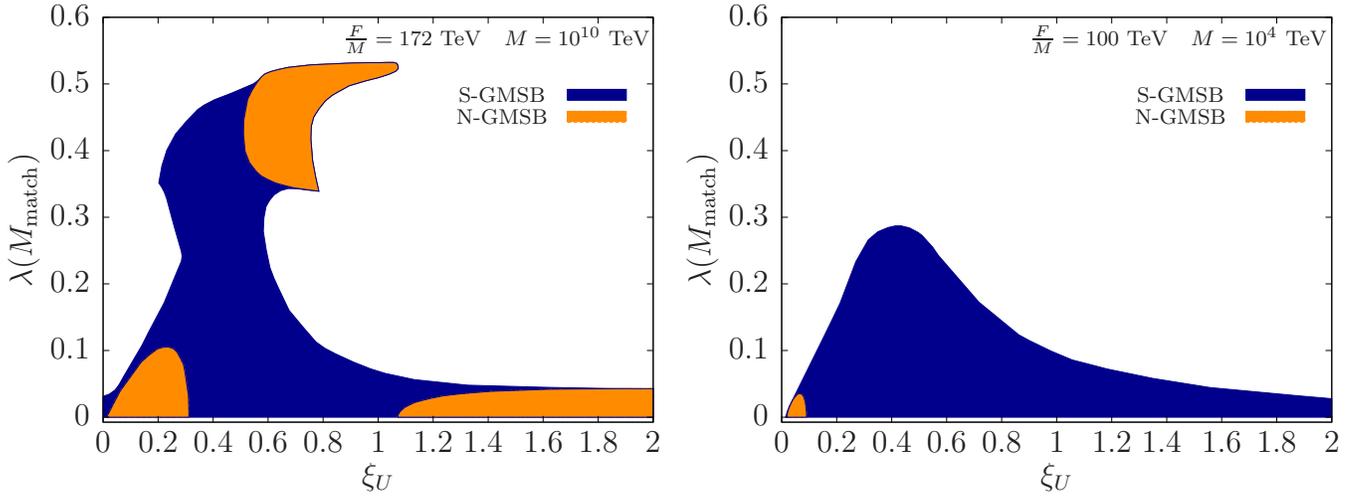}
\caption{Comparison of the allowed regions in the $\lambda(M_\mt{match})$-$\xi_U$ plane for the models with and without  $S \bar{\Phi}_1^D H_u$ superpotential interactions, S-GMSB and N-GMSB, respectively. Only the regions with $m_{H^1}\geq 114 \units{GeV}$ are shown. (Left) $F/M=172 \units{TeV}$ and $M=10^{10}\units{TeV}$ as in \cite{Delgado:2007rz}. (Right) The same for $F/M=100 \units{TeV}$ and $M=10^5\units{TeV}$. }
\label{fig_Comparison}
\end{figure}

In Fig. \ref{fig_Comparison} left we compare the allowed regions for both models in the $\lambda(M_\mt{match})$-$\xi_U$ plane for $F/M=172~\units{TeV}$ and $M=10^{10}\units{TeV}$. As it is apparent the $\xi_{H_u}$ coupling significantly increases the area where Higgs masses above 114 GeV can be attained, beyond the three characteristic regions explained in \cite{Delgado:2007rz}. This enhancement of the allowed regions is even more apparent when we move to lower SUSY-breaking scales, where the model in \cite{Delgado:2007rz} has problems in obtaining large enough Higgs masses. Indeed, in Fig. \ref{fig_Comparison} right we can observe how for $F/M=100\units{TeV}$, $M=10^4\units{TeV}$ the N-GMSB scenario is almost ruled out by the experimental constraints, while still a large region in the $\lambda(M_\mt{match})$-$\xi_U$ plane is allowed once we introduce the extra coupling. Let us also note that this does not require too large values for $\xi_{H_u}$. For the regions in Fig. \ref{fig_Comparison} we find $\xi_{H_u}\lesssim 4$, with $\xi_{H_u}\sim 4$ only at the end of the tail in Fig. \ref{fig_Comparison} right ($\xi_{H_u}\lesssim 1.5$ in Fig. \ref{fig_Comparison} left). Characteristic values for the sparticle masses in this latter example, and thus not easily accessible to the N-GMSB scenario, are the following: the lightest stop (sbottom) mass is $\sim 1.1$-$1.3\units{TeV}$ ($1.4$-$1.5\units{TeV}$); for the lightest stau $m_{\tilde{\tau_1}}\sim 220$-$270\units{GeV}$; for the gaugino masses $M_{\lambda_a}\approx 280, 510,  1400\units{GeV}$ for bino, winos and gluinos, respectively. Notice that the stau is lighter than the bino. The lightest neutralino is in general an admixture of Higgsinos and the bino, though regions where it is purely singlino are still present. In the latter case as well as in those regions where the effective $\mu$ term is low enough the lightest neutralino can still be the NLSP. Notice also that this is a short-lived NLSP ($\sqrt{F}=10^3\units{TeV}$), which might still decay within the detector at collider experiments giving a hard photon or two jets plus missing $E_T$, depending on whether the NLSP is mostly a bino or has a singlino/Higgsino component. In the case where the NSLP is a stau the decay inside the detector will provide a hard jet plus missing energy.

In summary, despite the model in \cite{Delgado:2007rz} is more predictive and it is a phenomenologically viable solution to the $\mu$-$b_\mu$ problem, at the price of only one more parameter the model presented here provides not only a significant enhancement of the allowed regions for those values of $F/M$ and $M$ where N-GMSB works but also extends the applicability of the gauge mediated NMSSM to scenarios where SUSY breaking occurs at lower energy scales.


\section{Conclusions}

In this paper we have extended the work presented in \cite{Delgado:2007rz}, where a model including singlet-messenger superpotential interactions was proposed in order to ameliorate the difficulties in generating the correct EWSB and a phenomenologically valid spectrum in gauge-mediation scenarios of the NMSSM. Such model thus also provides a satisfactory explanation to the $\mu$-$b_\mu$ problem of GMSB. However, it requires a relatively large scale of SUSY breaking in order to overcome all existing bounds and, in particular, the direct LEP 2 bound on the lightest CP-even Higgs boson mass. Considering an extension where we also include interactions between the NMSSM singlet, the messengers and the Higgs doublets we have proved that it is possible to lower significantly the scale of SUSY breaking without drastically constraining the size of the allowed regions in the parameter space. We have performed a scan over all the parameter space and discussed the general aspects of the spectrum of the model. Most of the basic features of a GMSB-like spectrum prevail in this scenario: strongly interacting particles are heavier, gravitino LSP, bino-like neutralino or stau NLSP, and so on. These are ``deflected'' by the new interactions, whose effects are multiplied by the NMSSM superpotential couplings of the corresponding particles. Thus, for instance, there is a significant splitting between the stops and first two families of up squarks that goes beyond the RG effects in GMSB. The overall scale of the spectrum can be relatively low. In particular, there are regions in the parameter space that allow stop masses below the TeV. Such values are not easily accessible without the new interactions. This low spectrum could be discovered at the early stages of the LHC.

We also emphasize that despite the general scenario introduces several extra free parameters, only one of them, the coupling $\xi_{H_u}$, is actually required in order to open the new regions in the parameter space. Thus, in practice, only one extra parameter is relevant and the model still remains quite predictive.

\section*{Acknowledgements}

It is a pleasure to thank G. F. Giudice and E. Ponton for reading the manuscript. We would like to especially thank P. Slavich for a lot of useful comments and clarifications. This work has been supported in part by the U.S. National Science Foundation under Grant PHY-0905283-ARRA.

\newpage

\appendix

\section*{Appendix: Renormalization Group Equations}
\label{app:}

In this appendix we provide formulae for the one-loop RG equations of the model presented in this paper. These can be derived using the general formulae in Refs. \cite{Martin:1993zk,Yamada:1993ga,Yamada:1994id}. Below the messenger scale $M$ these are simply given by the NMSSM ones, which can be found for instance in \cite{Ellwanger:2009dp}. It is important to note that, because of the contributions proportional to $\xi_{H_u,H_d}$ in the boundary conditions of the scalar soft masses squared, the boundary condition for the $U\left(1\right)_Y$ Fayet-Iliopoulos $D$ term
\be
\Delta \frac{d m_{\phi_i}^2}{d \log{Q}}= \frac{3}{5}\frac{g_1^2}{8\pi^2}~\!Y_{\phi_i}~\!\mathrm{Tr} \left[Y m^2_\phi\right]
\ee
is not vanishing at $M$. Therefore, this term contributes in the running unlike in the pure GMSB scenario or in the model presented in \cite{Delgado:2007rz}. Above the messenger scale the contributions from the messenger superpotential couplings modify the RG equations respect to those of the NMSSM. For the gauge couplings these are given by
\be
\begin{split}
\frac{d g_1}{d \log{Q}}=&\frac{g_1^3}{16\pi^2}\left(n+\frac{33}{5}\right),\\
\frac{d g_2}{d \log{Q}}=&\frac{g_2^3}{16\pi^2}\left(n+1\right),\\
\frac{d g_3}{d \log{Q}}=&\frac{g_3^3}{16\pi^2}\left(n-3\right),
\end{split}
\ee
where $n$ is the number of messengers and recall $g_1=\sqrt{\frac 5 3} g^\prime$. As explained in the main text we neglect any possible superpotential couplings between the messengers and the quark and lepton chiral fields at the messenger scale. However, as also emphasized there, even in that case nonvanishing values for the couplings with the third family are generated in the running above $M$. Thus, we must also include the effect of such couplings, as well as their RG equations in our analysis. For the NMSSM superpotential interactions we have

\be
\begin{split}
\frac{d y_t}{d \log{Q}}=&\frac{y_t}{16\pi^2}\left(6 y_t^2+y_b^2+\lambda^2+\xi_{H_u}^2+6\xi_t^2+\xi_b^2-\frac{13}{15}g_1^2-3g_2^2-\frac{16}{3}g_3^2\right)+\\
&+\frac{\xi_t}{16\pi^2}\left(\lambda \xi_{H_d} +\xi_D\xi_{H_u}\right),\\
\frac{d y_b}{d \log{Q}}=&\frac{y_b}{16\pi^2}\left(y_t^2+6y_b^2+y_\tau^2+\lambda^2+\xi_{H_d}^2+\xi_t^2+6\xi_b^2-\frac{7}{15}g_1^2-3g_2^2-\frac{16}{3}g_3^2\right)+\\
&+\frac{\xi_b}{16\pi^2}\left(y_\tau \xi_\tau+ \lambda \xi_{H_u}+\xi_D\xi_{H_d}\right),\\
\frac{d y_\tau}{d \log{Q}}=&\frac{y_\tau}{16\pi^2}\left(3y_b^2+4y_\tau^2+\lambda^2+\xi_{H_d}^2+4\xi_\tau^2-\frac{9}{5}g_1^2-3g_2^2\right)+\\
&+\frac{\xi_\tau}{16\pi^2}\left(3y_b \xi_b +\lambda \xi_{H_u} + \xi_D \xi_{H_d}\right),\\
\frac{d \lambda}{d \log{Q}}=&\frac{\lambda}{16\pi^2}\left(3y_t^2+3y_b^2+y_\tau^2+4\lambda^2+2\kappa^2+2\xi_D^2+3\xi_T^2+4\xi_{H_d}^2+4\xi_{H_u}^2-\frac{3}{5}g_1^2-3g_2^2\right)+\\
&+\frac{1}{16\pi^2}\left(3y_t \xi_{H_d} \xi_t+3y_b \xi_{H_u}\xi_b+y_\tau \xi_{H_u} \xi_\tau +2\xi_D\xi_{H_d}\xi_{H_u}\right),\\
\frac{d \kappa}{d \log{Q}}=&\frac{3\kappa}{16\pi^2}\left(2\lambda^2+2\kappa^2+2\xi_D^2+3\xi_T^2+2\xi_{H_d}^2+2\xi_{H_u}^2\right).\\
\end{split}
\ee

The RG equations for the  messenger superpotential couplings in Eq. (\ref{MessW}) read

\be
\begin{split}
\frac{d \xi_D}{d \log{Q}}=&\frac{\xi_D}{16\pi^2}\left(2\lambda^2+2\kappa^2+4\xi_{D}^2+3\xi_{T}^2+4\xi_{H_d}^2+4\xi_{H_u}^2+3\xi_t^2+3\xi_b^2+\xi_\tau^2-\frac 3 5 g_1^2-3g_2^2\right)+\\
&+\frac{1}{16\pi^2}\left(3y_t\xi_{H_u}\xi_t+3y_b\xi_{H_d}\xi_b+y_\tau\xi_{H_d}\xi_\tau+2\lambda\xi_{H_d}\xi_{H_u}\right),\\
\frac{d \xi_T}{d \log{Q}}=&\frac{\xi_T}{16\pi^2}\left(2\lambda^2+2\kappa^2+2\xi_{H_d}^2+2\xi_{H_u}^2+2\xi_D^2+5\xi_T^2-\frac{4}{15}g_1^2-\frac{16}{3}g_3^2\right),\\
\frac{d \xi_{H_d}}{d \log{Q}}=&\frac{\xi_{H_d}}{16\pi^2}\left(3y_b^2+y_\tau^2+4\lambda^2+2\kappa^2+4\xi_D^2+3\xi_T^2+4\xi_{H_d}^2+2\xi_{H_u}^2+3\xi_t^2-\frac35 g_1^2-3g_2^2\right)+\\
&+\frac{1}{16\pi^2}\left(3y_t\lambda\xi_t+3y_b\xi_D\xi_b+y_\tau\xi_D\xi_\tau+2\lambda\xi_D\xi_{H_u}\right),\\
\frac{d \xi_{H_u}}{d \log{Q}}=&\frac{\xi_{H_u}}{16\pi^2}\left(3y_t^2+4\lambda^2+2\kappa^2+4\xi_D^2+3\xi_T^2+2\xi_{H_d}^2+4\xi_{H_u}^2+3\xi_b^2+\xi_\tau^2-\frac35 g_1^2-3g_2^2\right)+\\
&+\frac{1}{16\pi^2}\left(3y_t\xi_D\xi_t+3y_b\lambda\xi_b+y_\tau\lambda\xi_\tau+2\lambda\xi_D\xi_{H_d}\right).
\end{split}
\label{MessCoupRGEs}
\ee

Finally, as explained above, we also need the RG equations for the couplings in Eq. (\ref{MessW2}) in the case of couplings with the third family only:

\be
\begin{split}
\frac{d \xi_t}{d \log{Q}}=&\frac{\xi_t}{16\pi^2}\left(6 y_t^2+y_b^2+\xi_D^2+\xi_{H_d}^2+6\xi_t^2+\xi_b^2-\frac{13}{15}g_1^2-3g_2^2-\frac{16}{3}g_3^2\right)+\\
&+\frac{y_t}{16\pi^2}\left(\lambda \xi_{H_d}+\xi_D\xi_{H_u}\right),\\
\frac{d \xi_b}{d \log{Q}}=&\frac{\xi_b}{16\pi^2}\left(y_t^2+6y_b^2+\xi_D^2+\xi_{H_u}^2+\xi_t^2+6\xi_b^2+\xi_\tau^2-\frac{7}{15}g_1^2-3g_2^2-\frac{16}{3}g_3^2\right)+\\
&+\frac{y_b}{16\pi^2}\left( y_\tau \xi_\tau+ \lambda \xi_{H_u}+\xi_D \xi_{H_d}\right),\\
\frac{d \xi_\tau}{d \log{Q}}=&\frac{\xi_\tau}{16\pi^2}\left(4y_\tau^2+\xi_D^2+\xi_{H_u}^2+3\xi_b^2+4\xi_\tau^2-\frac{9}{5}g_1^2-3g_2^2\right)+\\
&+\frac{y_\tau}{16\pi^2}\left(3y_b \xi_b +\lambda  \xi_{H_u}+\xi_D \xi_{H_d}\right).
\end{split}
\label{MessCoupRGEs2}
\ee
%



\end{document}